\pdfoutput=1
\documentclass{PoS}

\usepackage{amsmath}

\title{Lattice study for conformal windows of SU(2) and SU(3) gauge
  theories with fundamental fermions\thanks{Report numbers:HUPD-1507, LLNL-PROC-678652.}}

\ShortTitle{Conformal windows of SU(2) and SU(3) gauge theories}

\author{Cynthia Y.-H. Huang\\
        Institute of Physics, National Chiao-Tung University, Hsinchu
        30010, Taiwan\\
        E-mail: \email{cynthia1122.py01g@nctu.edu.tw}}

\author{Issaku Kanamori\\
        Institute of Physics, National Chiao-Tung University, Hsinchu
        30010, Taiwan\\
        Graduate School of Science, Hiroshima University,
        Higashi-Hiroshima, Japan\thanks{Current address.}\\
        E-mail: \email{kanamori@hiroshima-u.ac.jp}}

\author{\speaker{C.-J. David Lin}\\
        Institute of Physics, National Chiao-Tung University, Hsinchu
        30010, Taiwan\\
        E-mail: \email{dlin@mail.nctu.edu.tw}}

\author{Kenji Ogawa\\
        Institute of Physics, National Chiao-Tung University, Hsinchu
        30010, Taiwan\\
        E-mail: \email{ogawaknj@gmail.com}}

\author{Hiroshi Ohki\\
        RIKEN-BNL Research Centre, Brookhaven National Laboratory,
        Upton, NY 11973, USA\\
        E-mail: \email{hoki@quark.phy.bnl.gov}}

\author{Alberto Ramos\\
        PH-TH, CERN CH-1121 Geneva 23, Switzerland\\
        E-mail: \email{alberto.ramos@cern.ch}}

\author{Enrico Rinaldi\\
        Lawrence Livermore National Laboratory, Livermore, CA 94550, USA\\
        E-mail: \email{rinaldi2@llnl.gov}}

\abstract{We present our investigation of SU(2) gauge theory with
  8 flavours, and SU(3) gauge theory with 12 flavours.  For
  the SU(2) case, at strong bare coupling, $\beta \lesssim
  1.45$, the distribution of the lowest eigenvalue of the Dirac
  operator can be described by chiral random matrix theory for the
  Gaussian symplectic ensemble.
 Our preliminary result indicates that the chiral phase
  transition in this theory is of bulk nature.  For the SU(3) theory,
we use high-precision lattice data to perform the step-scaling study
of the coupling,
$g_{_{{\rm GF}}}$, in the
Gradient Flow scheme.  
We carefully examine the reliability of the
continuum extrapolation in the analysis, 
and conclude that the scaling behaviour
of this SU(3) theory is not governed by possible infrared conformality at
$g_{_{{\rm GF}}}^{2} \lesssim 6$.}

\FullConference{The 33rd International Symposium on Lattice Field Theory\\
		14 -18 July 2015\\
		Kobe International Conference Center, Kobe, Japan}

\begin{document}

\section{Introduction}
\label{sec:intro}
The search for infrared (IR) conformality in various
gauge theories has recently been a popular subject in Lattice Field
Theory~\cite{Lucini:2015noa,DeGrand:2015zxa}.   Results of such research
activities can lead to useful information for constructing
composite Higgs, or walking technicolour (WTC), models.  These models contain dynamical electroweak
symmetry breaking, and have been shown to be compatible with
experiments at the LHC~\cite{Elander:2012fk,Matsuzaki:2012mk}.
Given a gauge group and a representation of the
fermions, it is essential to determine the smallest number of
flavours, $N_{f}^{{\rm cr}}$,
where the theory can contain an infrared fixed point (IRFP).  The
value of $N_{f}^{{\rm cr}}$ is often referred to as the lower end of
the conformal window for a family of theories.  A
viable WTC model can be obtained with the number of flavours just
below $N_{f}^{{\rm cr}}$, such that the infrared (IR) nearly scale-invariant
behaviour of the theory can produce a state that is
parametrically light~\cite{Dietrich:2005jn, Appelquist:2010gy}.

In lattice
determination of $N_{f}^{{\rm cr}}$, the need for accurate numerical
calculations results from the
challenge in distinguishing between IR conformality and
slow-running behaviour in theories near the lower end of the conformal
window.  One example is the 12-flavour SU(3) gauge theory that has
been studied by many groups.  The majority
concluded that the theory is IR conformal.  Amongst these investigations,
one popular strategy is the step-scaling method for computing the
running coupling.   This method was implemented using the
Schr\"{o}dinger Functional (SF) scheme~\cite{Appelquist:2009ty} and
the Twisted Polyakov Loop (TPL)
scheme~\cite{Lin:2012iw,Itou:2012qn}, with a variation of it carried out
for the Gradient Flow (GF) scheme ~\cite{Cheng:2014jba}.   All these
previous studies lead to evidence for IR
scale-invariance.  Nevertheless, the smallness of the
$\beta$-function in this theory has to be noted.
Two-loop perturbation theory predicts at most $6\%$ change of the
renormalised coupling between the Gaussian ultraviolet (UV) and the possible
strongly-coupled IR fixed points at doubling the length scale.
Therefore, to make any statistically-meaningful statement regarding
the running behaviour, it is desirable to have results with
subpercentage-level error.  Such precision was not achieved in
the previous works, leaving room to improve the calculations.  In
this article we present our result of the step-scaling study
for the GF-scheme coupling in this theory.  We
obtain the renormalised coupling at $\lesssim 0.5\%$  statistical
error, and use two discretisations which allow us
to check the reliability of the continuum extrapolation.

In this presentation, we also show preliminary results from our investigation of
the chiral phase transition in SU(2) gauge theory with 8 flavours.  As pointed out in
Ref.~\cite{Appelquist:2013pqa}, the 6-flavour SU(2) gauge theory can
be confining.  This makes the study of the 8-flavour theory interesting
for determining $N_{f}^{{\rm cr}}$ for the case of the SU(2) gauge
group.   This theory may be IR conformal or confining with a very small
$\beta$-function as well.  There exist relatively few
results concerning its IR
behaviour~\cite{Ohki:2010sr,Rantaharju:2015yva}, and our investigation
can lead to further information in this regard.
In particular, we
compute the distribution of the lowest eigenvalue of the Dirac
operator, and compare with the prediction from the Random Matrix
Theory (RMT).   This method
can be used to extract the infinite-volume chiral condensate from
finite-volume lattice data, and there is no contamination
from power divergences.  Therefore it leads to more reliable results.
 In addition to its application in QCD, this
approach has also been adopted for lattice computations for beyond the
standard model physics in Ref.~\cite{Fodor:2009wk}.   Using this
strategy, we find the existence of the chirally broken phase in SU(2)
gauge theory with 8 flavours.  Our preliminary results show that the
chiral phase transition is of bulk nature.

\section{Random Matrix Theory and SU(2) gauge theory with eight flavours}
\label{sec:rmt_su2}
We use the plaquette gauge action
and unimproved staggered fermions in the simulation.
The lattice volumes, $\hat{V}=\hat{T}\times\hat{L}^{3}$, are
$\hat{T}=\hat{L} =8$, $12$ and $16$.
The fermion masses are $\hat{m}_{f} = am_f=0.005$, $0.010$ and $0.015$.
We study the phase structure of this system by computing
the chiral condensate.
Investigation of the plaquette and Polyakov loop of this theory was reported
in \cite{Huang:2014xwa}.
We use chiral RMT to extract the chiral
condensate from our data.  The RMT provides a reliable procedure
which is free from power divergences and finite volume
effects, if the system is in the $\epsilon$-regime.

The dynamical variables for the RMT are the eigenvalues of an $N\times N$
matrix, where $N$ should be taken $N\to\infty$.
After suitable rescaling, the eigenvalues $\zeta_i$ follow
the distribution:
\begin{equation}
 \rho_N^{(\beta)}(\zeta_1,\dots,\zeta_N; \mu_1,\dots,\mu_{N_f})
=
 C
  \prod_{i=1}^N
  \left(
   \zeta_i^{\beta(\nu+1)-1} e^{-\beta \frac{\zeta_i^2}{8N}}
   \prod_{a=1}^{N_f}(\zeta_i+\mu_a^2)
 \right)
  \prod_{i>j}^N|\zeta_i^2-\zeta_j^2|^\beta,
  \label{eq:dist-zeta}
\end{equation}
where $N_f$ is the number of flavours,
 $\mu_a$ are mass parameters, $\nu$ is the topological charge, and $C$
is the overall normalisation.
The Dyson index $\beta$ is $4$ for staggered fermions in the SU(2)
fundamental representation~\cite{Damgaard:2001fg}.
As pointed out in \cite{BerbenniBitsch:1998sy},
the corresponding number of flavours for RMT
is different from that for the lattice simulation $N_f^{\mathrm{lat.}}$:
\begin{equation}
 N_f = 2\cdot \frac{1}{4} \cdot N_f^{\mathrm{lat.}} = 4,
\end{equation}
where the factor $2$ comes form the 2-fold degeneracy due to pseudo
reality of the SU(2) gauge group, 
and $1/4$ is from the taste breaking of staggered fermion.

In this work, we use the distribution of the lowest eigenvalue
derived from Eq.~(\ref{eq:dist-zeta})~\cite{Damgaard:2000ah}.
The lowest eigenvalue, $\lambda_1$, of the massless Dirac operator 
computed on the lattice should follow the same distribution, after rescaling with the
chiral condensate $\Sigma$, as predicted by the RMT:
\begin{equation}
 p_1^{\mathrm{RMT}}(\zeta_1; \mu)
\Big|_{\zeta_1=\lambda_1 V\Sigma, \mu=m_f V\Sigma} 
= \frac{1}{\hat{V}\hat{\Sigma}}p_1^{\mathrm{latt.}}(\hat{\lambda}_1; \hat{m}_f) , 
\label{eq:rmt-lattice}
\end{equation}
where $\hat{\Sigma} = a^{3} \Sigma$, 
$\hat{\lambda}_{1} = a \lambda_{1}$, 
$V=a^{4} \hat{V}$ and $m_f = \hat{m}_{f}/a$, with $\hat{V}$ and $\hat{m}_{f}$
being inputs in our lattice
simulations.
Equation~(\ref{eq:rmt-lattice}) can be used to determine
$\hat{\Sigma}$, 
given the $p_{1}^{{\rm latt}} (\lambda_{1};m_{f})$ computed on the
lattice, and the RMT-predicted $p_1^{\mathrm{RMT}}(\zeta_1,\mu)$.
Non-applicability of the RMT for extracting the condensate
means the restoration of chiral symmetry.
The analytic expression of $p_1^{\mathrm{RMT}}(\zeta_1,\mu)$
can be complicated.  In fact, the available analytic result for
the Gaussian symplectic ensemble is only applicable to cases of
fractional topological charges~\cite{Damgaard:2000ah}.
For this reason, we use the Hybrid Monte Carlo (HMC) method to obtain the distribution
with $\zeta_i$ treated as the dynamical variables in Eq.~(\ref{eq:dist-zeta}).
We find that $N=400$ is large enough, such that our results do not show significant
$N$-dependence.
Distributions with various values of $\mu$ from $0.0$
to $100.0$ are obtained by the HMC strategy, and then interpolated to $\mu= m_f V\Sigma$.

\begin{figure}[t]
\noindent
\centering
 \includegraphics[width=0.49\linewidth]{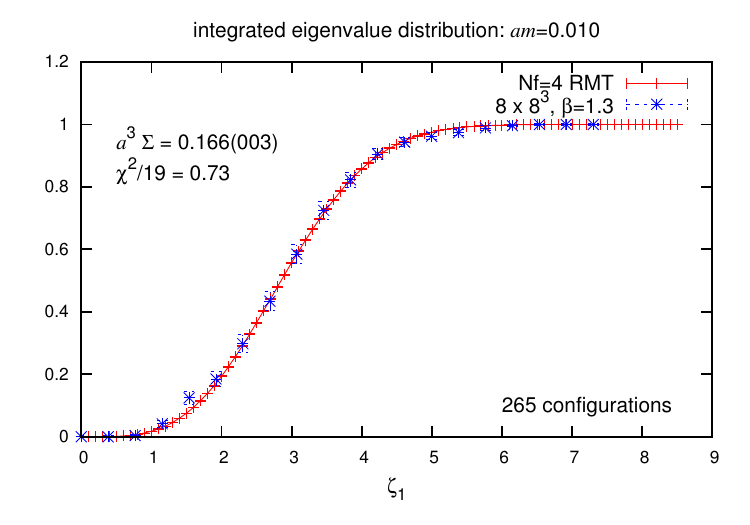}
 \hfil
  \includegraphics[width=0.49\linewidth]{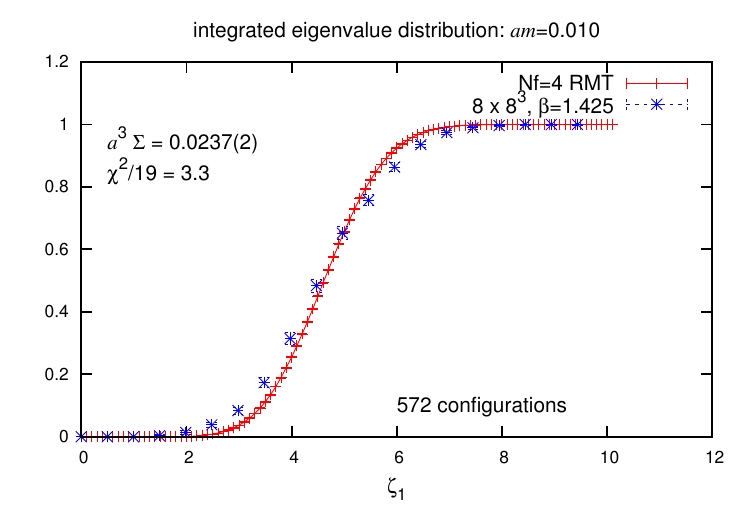}
 \caption{Examples of good fit (left) and bad fit (right) of the chiral
 condensate $\Sigma$ by using the integrated distribution of lowest
 eigenvalue at $\nu = 0$.  The red lines are from $N_f=4$ RMT,
 and the blue crosses are the lattice
 data, with the eigenvalue, $\lambda_{1}$,  rescaled to 
 $\zeta_1 = V\Sigma\lambda_1$ using the value of $\Sigma$ shown in the plot.}
 \label{fig:good-and-bad}
\end{figure}

Two typical examples of the fits at $\nu = 0$ are shown in
Fig.~\ref{fig:good-and-bad}.
The bad fit (right panel, $\beta = 1.475$) gives a very small value of
$\Sigma$, indicating the restoration of chiral symmetry\footnote{%
We also observed that $N_f=8$ and $N_f=16$ RMT have a tendency to give
poorer fitting, which is consistent with the taste breaking.
Our data does not show clustering of the eigenvalues for the
taste symmetry.}.
According to the previous study of this system \cite{Huang:2014xwa},
the theory at $\beta=1.475$ 
is in the weak-coupling phase, while at $\beta=1.3$ 
it is in the strong-coupling phase.

\begin{figure}[t]
 \noindent
 \centering
 \includegraphics[width=0.45\linewidth]{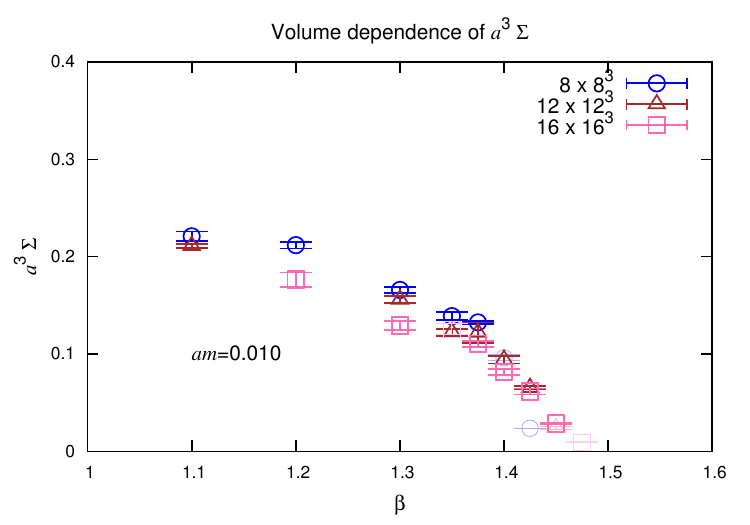}
 \hfil
 \includegraphics[width=0.45\linewidth]{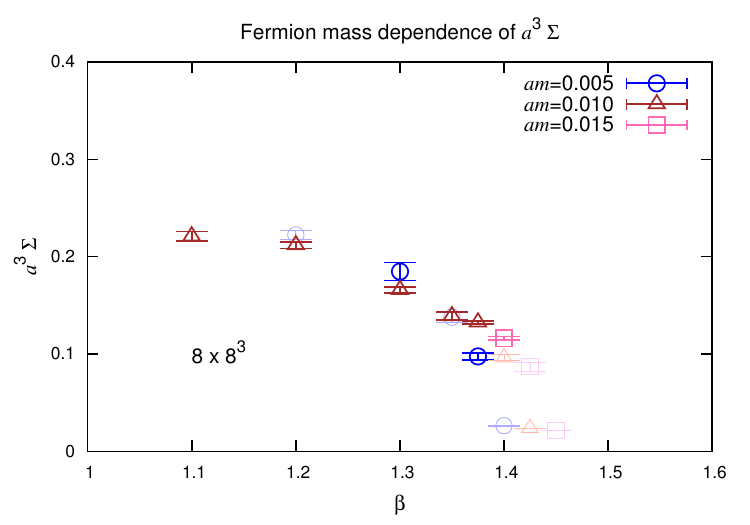}
  \noindent
 \centering
 \includegraphics[width=0.45\linewidth]{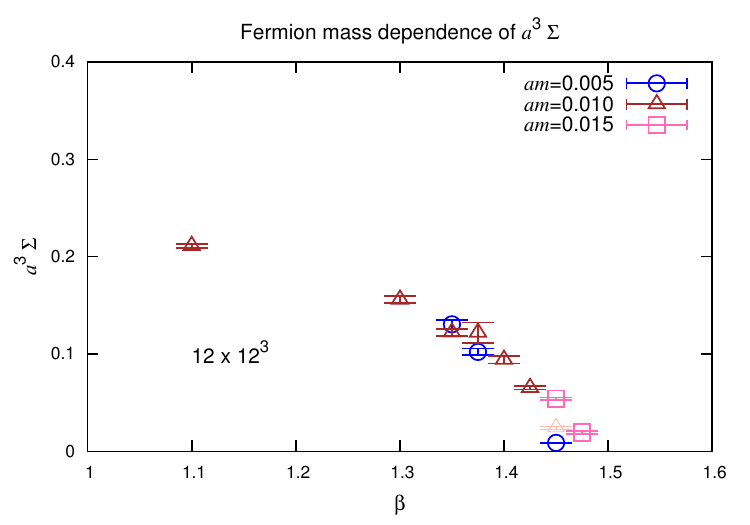}
 \hfil
 \includegraphics[width=0.45\linewidth]{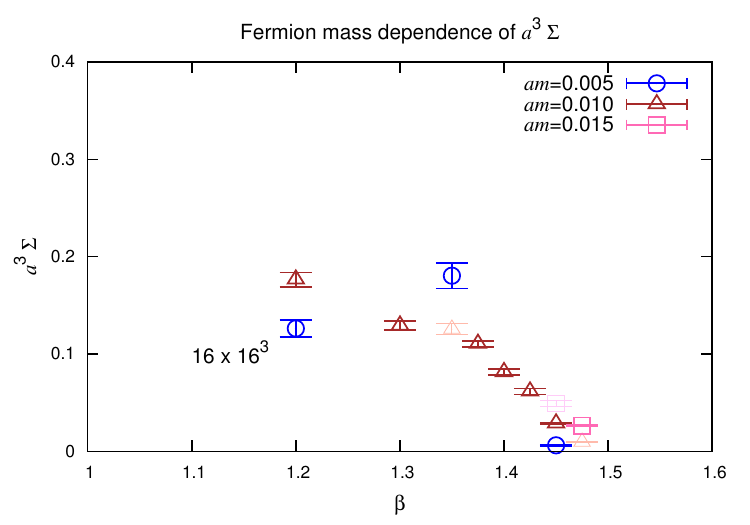}

 \caption{Chiral condensate $\Sigma$, extracted in the $\nu = 0$
   sector, versus the bare coupling $\beta$.
 The upper left panel is with fermion mass $am_f=0.010$ and several lattice
 volumes.  The other plots are at fixed lattice volume with several fermion mass.
 Thin symbols indicate poor fitting of $\Sigma$ with
 $\chi^2/{\mathrm{d.o.f}}\geq 1$.
 }
 \label{fig:sigma}
\end{figure}

Figure~\ref{fig:sigma} displays plots of $\Sigma$ against $\beta$.
All the panels show that the chiral condensate vanishes
for $\beta \gtrsim 1.45$, and most of the fit results
with small $\Sigma$ come with poor $\chi^2/{\mathrm{d.o.f}}$ (thin symbols).
The location of this chiral phase transition is observed to have 
almost no volume (upper-left panel) and fermion-mass (other panels) dependence.
This implies that the transition is of bulk nature,
and the phase connecting to the weak-coupling continuum limit is the
symmetric phase\footnote{Note that the continuum 8-flavour SU(2)
  theory is described by the RMT for the Gaussian orthogonal ensemble~\cite{Damgaard:2001fg},
and we are aware that this may complicate the interpretation of our results.}.
We are currently generating lattice data at larger volumes to further
check the validity of this scenario.  More detailed analysis of systematic errors,
such as the effect of the taste breaking, is also needed.

\section{The Yang-Mills gradient flow and SU(3) gauge theory with
  twelve flavours}
\label{sec:GF_su3}
For the study of SU(3) gauge theory with 12 flavours, we adopt the
step-scaling method to investigate 
the renormalised running coupling, $g_{{\rm GF}}$, in the Gradient Flow (GF)
scheme~\cite{Luscher:2010iy,Fodor:2012td}.   In this work we use two discretisations, namely the
clover and the plaquette, for extracting this coupling.  This
allows us to examine the reliability of our results in the continuum limit.
To ensure that we only have one length scale for probing
the theory, it is necessary to fix the ratio $c_{\tau} = \sqrt{8
  t}/L$, where $L$ is the lattice size and $t$ is the flow time.  We implement the
colour-twisted boundary condition, and perform the simulations at
vanishing fermion mass.
The work presented here has been reported in our recent paper, Ref.~\cite{Lin:2015zpa}, which we refer
to for more details and unexplained notation. Our
goal is to compute, in the continuum limit,
\begin{equation}
\label{eq:r_sigma_def}
  r_{\sigma}(L) \equiv \frac{g^{2}_{{\rm GF}}(2 L)}{g^{2}_{{\rm GF}}(L)} ,
\end{equation}
where $L$ is interpreted as the
renormalisation scale.   To proceed, we first specify a value, $u$,
and tune the bare couplings on the
$\hat{L} \equiv L/a =8, 10, 12$ lattices, such that the renormalised couplings,
$\bar{g}^{2}_{{\rm latt}}$,
extracted on these lattices all match this value.  This $u$ is
interpreted as the continuum coupling, $g^{2}_{{\rm
    GF}}$, renormalised at $L$.
The details of this
tuning procedure is explained in Refs.~\cite{Lin:2012iw, Lin:2015zpa}.
We then use the bare couplings obtained above to determine the
lattice step-scaling functions, $\Sigma (u, a)$, which are simply
$\bar{g}^{2}_{{\rm latt}}$ computed on the corresponding lattices $2 \hat{L} =
16, 20, 24$.  This allows us to perform
the linear continuum extrapolation using the ans\"{a}tz {\it
  a'la} Symanzik, $ \Sigma (u, a) = \sigma (u) + A (a/L)^{2}$.
The ratio, $r_{\sigma}$, defined in Eq.~(\ref{eq:r_sigma_def}) can
then be expressed as $r_{\sigma} (u) = \sigma(u) / u$.

\begin{figure}[t]
\noindent
\centering
 \includegraphics[width=0.43\linewidth]{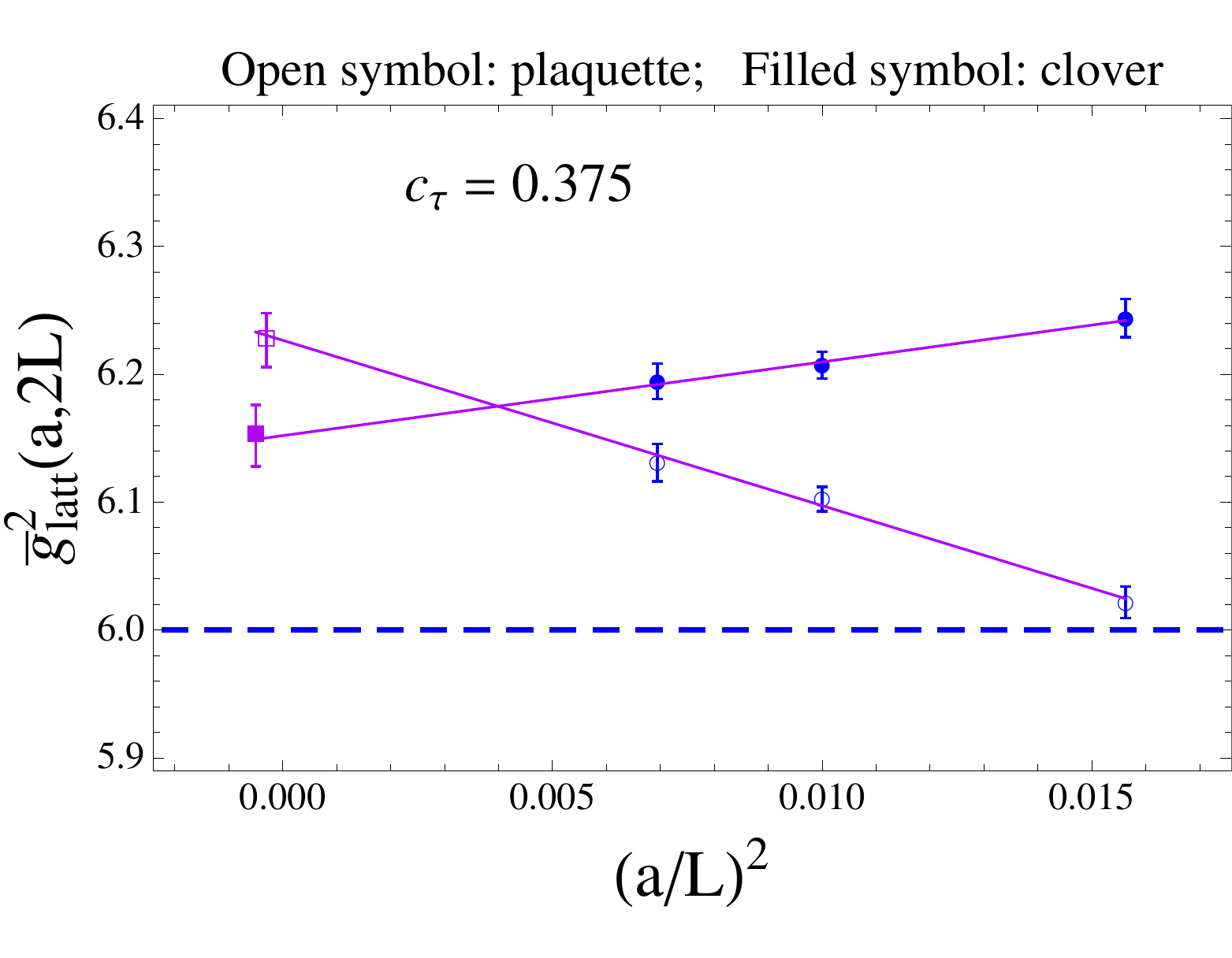}
 \hfil
  \includegraphics[width=0.43\linewidth]{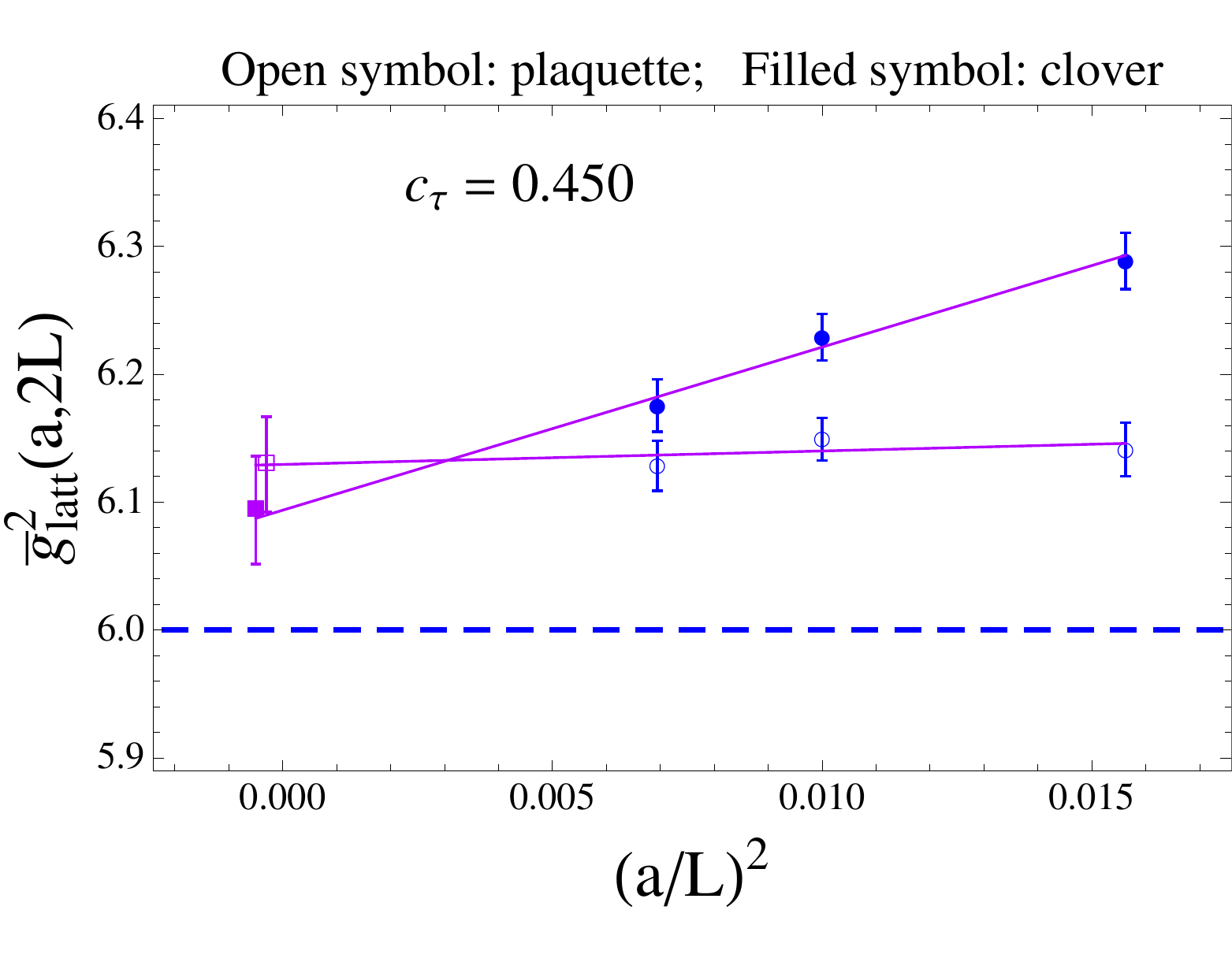}
  \caption{Continuum extrapolation for the step-scaling function for
    input $u=6.0$ at $c_{\tau} = 0.375$ and $0.45$.}
 \label{fig:cont_extrap_c_0.375_0.45}
\end{figure}

Figure~\ref{fig:cont_extrap_c_0.375_0.45} shows the continuum
extrapolation of the step-scaling function using 
the functional form linear in $(a/L)^{2}$, at $c_{\tau} = 0.375$ and
$0.45$, in the strong-coupling regime (input $g^{2}_{{\rm GF}} = 6$).
These plots clearly demonstrate that one has to be careful when
performing this extrapolation.  As can be seen for the case of
$c_{\tau} = 0.375$, the extrapolations are mild and smooth.  On the
other hand, the two discretisations do not lead to compatible result in
the continuum limit, indicating that the effects of the lattice
artefacts are significant.  The situation is improved at increasing $c_{\tau}$.  In
this work, we find that the continuum extrapolation is under control
at $c_{\tau} \ge 0.45$.

\begin{figure}[t]
\noindent
\centering
 \includegraphics[width=0.42\linewidth]{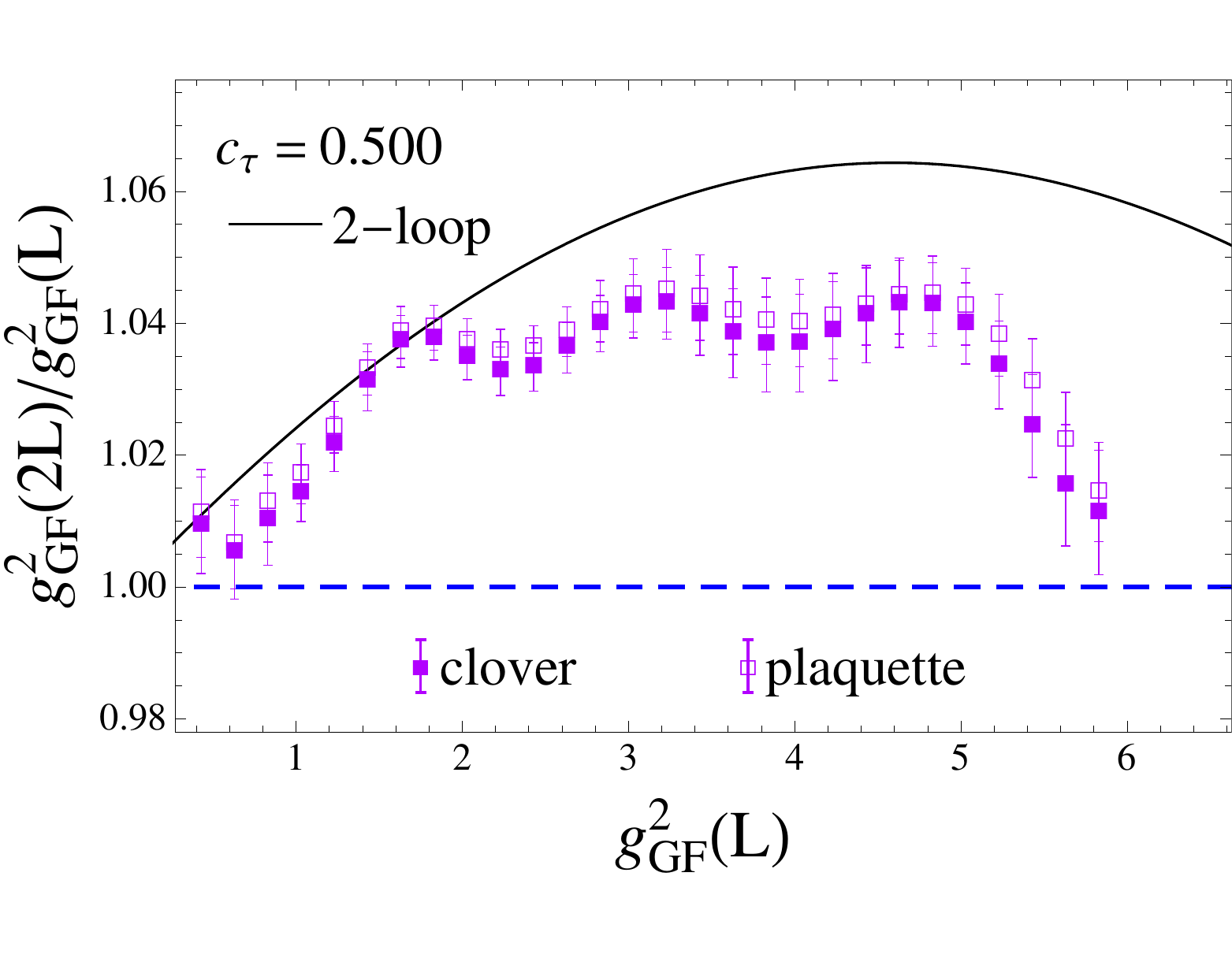}
 \hfil
  \includegraphics[width=0.43\linewidth]{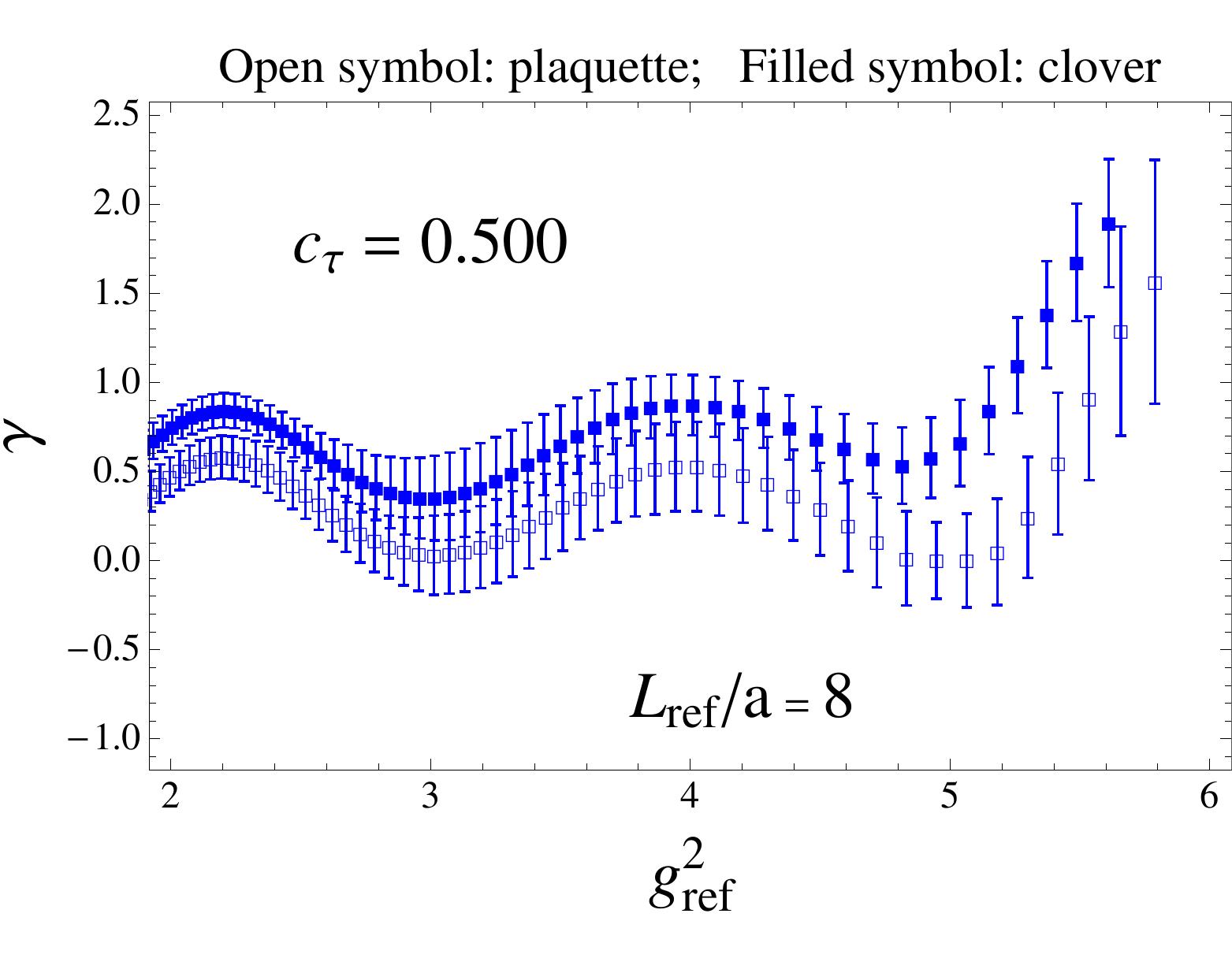}
  \caption{Left: Results of $r_{\sigma}$ plotted against input
    $g^{2}_{{\rm GF}}(L)$ at $c_{\tau} = 0.5$. Right: Plot of
    $\gamma$ against $g_{{\rm ref}}^{2}$.}
 \label{fig:r_sigma_c_0.375_0.5}
\end{figure}

Results of the ratio $r_{\sigma}$ for $c_{\tau} = 0.5$ are
displayed in the left panel Fig.~\ref{fig:r_sigma_c_0.375_0.5}.
From this plot, we observe that at $g^{2}_{{\rm GF}} \gtrsim 5$, the
GF-scheme coupling runs significantly slower than the two-loop
perturbative prediction.  At input $g^{2}_{{\rm GF}}(L) = 5.8$, $r_{\sigma}$ is almost
consistent with unity.   However,
one has to be cautious in using this result to indicate the existence
of an IRFP in this theory.  In Ref.~\cite{Lin:2015zpa}, we argue that the
continuum extrapolation of the lattice data for the coupling constant
near an IRFP can be a subtle issue, and it is desirable to perform a
finite-size scaling test.  This test involves fitting the
lattice numerical results for the (cutoff-dependent) renormalised coupling,
$\bar{g}_{{\rm latt}}^{2} (g_{0}^{2}, \hat{L} = L/a)$, at various values
of $\hat{L}$ to the formula
\begin{equation}
\label{eq:FSS}
 \bar{g}_{{\rm latt}}^{2}  (g_{0}^{2}, \hat{L}) = g^{2}_{l} (g_{{\rm
     ref}}) + \left [ g^{2}_{{\rm ref}} - g^{2}_{l} (g_{{\rm
       ref}})\right ] \left ( \frac{\hat{L}_{{\rm ref}} }{\hat{L}}
 \right )^{\gamma (g_{{\rm ref}})} ,
\end{equation}
where $g_{l}$ and $\gamma$ are free parameters, $g_{0}$ is the bare
coupling, $g_{{\rm ref}} = \bar{g}_{{\rm latt}}  (g_{0}^{2},
\hat{L} = \hat{L}_{{\rm ref}} = 8)$.  This formula is derived from the ``locally
linearised'' $\beta$-function.  This linearisation is valid (away
from the asymptotic-freedom limit) because
the $\beta$-function is small.   We implement the scaling test using
Eq.~(\ref{eq:FSS}) and scanning through many values of $g_{0}$.  At
each choice of $g_{0}$ (hence $g_{{\rm ref}}$ since we fix $\hat{L}=8$), we perform a fit.  When the bare coupling is
tuned such that the theory is in the vicinity of the possible IRFP,
$g_{l}$ and $\gamma$ will approach $g_{\ast}$ (the location of the
fixed point) and $\gamma_{\ast}$ (the slope of the $\beta$-function
at the zero in the strong-coupling regime).  That is, the signal for
the existence of IR conformality should be the plateaus in the plots
of $g_{l}$ and $\gamma$ against $g_{{\rm ref}}$.   In addition, scale
invariance should ensure that results from different discretisations
agree.  The outcome of this
analysis is shown in the right panel of
Fig.~\ref{fig:r_sigma_c_0.375_0.5}.   It is clear that our data do not
indicate the existence of an IRFP in the regime where our studies are
carried out.

\section{Conclusion and outlook}
\label{sec:conclusion}
In this talk, we present results from our lattice investigations for the IR
behaviour of SU(2) gauge theory with 8 flavours, and SU(3) with 12
flavours.  For the SU(2) case, our preliminary finding is that there
exists a chirally broken phase that can be well described by the RMT
of the Gaussian symplectic ensemble.  We find evidence that the
relevant chiral phase transition can be of bulk nature, and the theory
in the continuum limit may be chirally symmetric.  Presently we
are performing simulations at larger volumes to obtain further
information regarding this transition.  As for the
SU(3) theory, our study of the GF-scheme coupling shows that the
behaviour theory is not governed by possible IR conformality at
$g^{2}_{{\rm GF}} \lesssim 6$.\\ \\
{\bf \large Acknowledgments}
We are indebted to Tatsumi Aoyama and Hideo Matsufuru for their help
in developping the HMC code, and thank Luigi~Del~Debbio, Shinsuke~M.~Nishigaki,  and
Ben~Svetitsky for helpful discussion.  I.K. and C.-J.D.L. acknowledge
research grant 102-2112-M-009-002-MY3 from Taiwanese MOST.  The work of H.O. is supported by
   the RIKEN Special Postdoctoral Researcher program.  The JSPS Grant-in-Aid for Scientific Research
   for Young Scientists (B) No.25800139 is also acknowledged by H.O..
   E.R. acknowledges the support from the DOE under contract
   DE-AC52-07NA27344 (LLNL).
   Most of the computational work was performed at Taiwanese NCHC,
   and KMI ($\phi$) at Nagoya University.
C.-J.D.L. thanks the hospitality of CERN and Hiroshima University
during the progress of these projects.

\end{document}